\documentclass[10pt]{wlscirep}

\usepackage{capt-of}
\usepackage{amsmath, amssymb, physics}
\usepackage{graphicx}% Include figure files
\usepackage{dcolumn}% Align table columns on decimal point
\usepackage{bm}% bold math
\AtBeginDocument{\usepackage{booktabs}}

\begin{document}

\title{Fabrication of ultrahigh-precision hemispherical mirrors for quantum-optics applications}

\author[1,2,*]{Daniel B. Higginbottom}
\author[1]{Geoff T. Campbell}
\author[2]{Gabriel Araneda}
\author[3]{Fengzhou Fang}
\author[2]{Yves Colombe}
\author[1]{Ben C. Buchler}
\author[1,$\dag$]{Ping Koy Lam}

\affil[1]{
Centre for Quantum Computation and Communication Technology, Research School of Physics and Engineering, The Australian National University, Canberra ACT 2601, Australia.}
\affil[2]{
 Institut f\"{u}r Experimentalphysik, Universit\"{a}t Innsbruck, Technikerstr. 25, 6020 Innsbruck, Austria}
\affil[3]{
State Key Laboratory of Precision Measuring Technology \& Instruments, Centre of MicroNano Manufacturing Technology, Tianjin University, Tianjin 300072, China
}
\affil[*]{daniel.higginbottom@anu.edu.au}
\affil[$\dag$]{ping.lam@anu.edu.au}

\date{\today}

\begin{abstract}
High precision, high numerical aperture mirrors are desirable for mediating strong atom-light coupling in quantum optics applications and can also serve as important reference surfaces for optical metrology. In this work we demonstrate the fabrication of highly-precise hemispheric mirrors with numerical aperture NA = 0.996. The mirrors were fabricated from aluminum by single-point diamond turning using a stable ultra-precision lathe calibrated with an in-situ white-light interferometer.  Our mirrors have a diameter of 25~mm and were characterized using a combination of wide-angle single-shot and small-angle stitched multi-shot interferometry. The measurements show root-mean-square (RMS) form errors consistently below 25~nm. The smoothest of our mirrors has a RMS error of 14~nm and a peak-to-valley (PV) error of 88~nm, which corresponds to a form accuracy of $\lambda/50$ for visible optics.
\end{abstract}

%\pacs{42.79.-e, 87.85.Va, 81.16.-c, 32.80.Qk, 42.50.Lc}
%42.79.-e Optical elements, devices and systems
% 87.85.Va	Micromachining
%81.16.-c	Methods of micro- and nanofabrication and processing
% 32.80.Qk	Coherent control of atomic interactions with photons
%42.50.Lc	Quantum fluctuations, quantum noise, and quantum jumps

\flushbottom
\maketitle
\thispagestyle{empty}

\section*{Introduction}

The controlled interaction of optical fields with material quantum systems is an essential engineering requirement of quantum networks and sensors. Optical resonators enhance the interaction strength between light fields and emitters by tuning the optical mode density in the vicinity of the emitter \cite{E.M.Purcell1946, Kleppner1981a, Goy1983}, an observation that spurred seminal results in the field of cavity quantum electro-dynamics and enabled the successful use of resonator-emitter systems as photon sources, quantum network nodes and sensors \cite{Cirac1997,Vahala2003, Reiserer2015}. To achieve efficient coupling without modifying the free-space vacuum mode density it is necessary to match incoming light fields to an emitter's natural radiation pattern \cite{Leuchs2013}, time reversing the emission process to engineer strong absorption. For a point source emitting spherical waves the ideal mode converter is a deep parabolic mirror \cite{Sondermann2007, Lindlein2007}. This single-pass approach requires high numerical aperture (NA) reflectors with sub-wavelength surface precision \cite{Maiwald2012a}. However the fabrication of high-NA mirrors with sufficient surface smoothness and form precision  remains a technological challenge \cite{Kim2009, Kim2004, Leuchs2008}.

Whereas a parabolic mirror is the desired reflector for converting spherical waves into plane waves, a hemispheric mirror maximizes the self-interaction of a source by returning spherical waves from the source to their origin. The hemispheric mirror is an intriguing special case for high-NA atom-light couplers that bridges the gap between single-pass optics and cavity quantum electrodynamics. Like an optical resonator, a single hemispheric mirror may enhance atom-light interactions by shaping the vacuum mode density around an emitter, but unlike an optical resonator the hemisphere-mediated atom-light interaction is single-pass. It has been predicted \cite{Hetet2010} that the spontaneous emission rate of an atomic electron at the center of curvature (CoC) of a spherical mirror may be suppressed or enhanced depending on the radius of the mirror, even when the mirror radius is much larger than the atomic wavelength. A previous attempt to measure such an effect with spherical optics measured emission rate fluctuations of 1\% \cite{Eschner2001}. An ideal hemisphere that covers exactly half of the atomic emission solid angle will achieve the greatest possible modification, enhancing the spontaneous emission rate by a factor of two when the radius is $R= n \frac{\lambda}{2} + \frac{\lambda}{4}$ where n is a positive integer and $\lambda$ is the transition wavelength, and completely suppressing spontaneous emission when the radius is $R = n \frac{\lambda}{2}$. Because deviations from an ideal hemisphere reduce the degree of suppression and enhancement, demonstrating this effect requires the fabrication of hemispheric mirrors with great surface precision. 

The fabrication of ideal reference optics is also of interest to the broader optics community. The hemisphere retro-reflects an incoming spherical wavefront and could be used as a reference null for characterizing high-NA focusing optics \cite{Inoue2003}. Optical-reference microscopy requires reference surfaces considerably better than the surface to be measured. The roundest manufactured objects are convex spheres made for NASA's Gravity Probe B rotors, which deviate by $17$ nm from perfect spheres (peak to valley)\cite{Turneaure1989, Everitt2011}, closely followed by the Avogadro spheres, which are $89$ nm from perfect \cite{ Leistner1991, Andreas2011a}. We understand from personal correspondence with the NMI and from the science media\cite{eveleth2014} that the Avogadro project has since improved their spheres to within $25$ nm (peak to valley) of perfect. Such outstandingly spherical spheres are produced by randomly rotating the sphere between two conical grinding/polishing tools for periods of several days. It is difficult to polish concave surfaces to comparable precision because no equivalent symmetric mounting process is possible, and there is an unmet demand for reference optics in this regime.

We diamond turn a nm-precise hemispheric mirror from an aluminum substrate and describe the challenges encountered when cutting such high NA concave optics. The mirror surface quality is verified by complementary measurements including single-shot optical interferometry with a reference sphere, multi-shot interferometry with a reference flat and contact probe measurements. We demonstrate the capacity to cut concave hemispheres that surpass the requirements for single-pass QED experiments, and characterize five consecutively manufactured mirrors as a test of consistency and reproduceability.

\section*{Materials and methods}\label{methods}

Single point diamond turning (SPDT) is an established tool for manufacturing ultra-fine optics with geometries accurate below optical wavelengths and smooth surface finishes. With SPDT it is possible to achieve both high material removal rates and low subsurface damage, making it an appealing technique for generating spheric and aspheric surfaces as well as rotationally asymmetric free-form elements for telescopes and head-up displays \cite{JannickP.Rolland, Kim2004, Leuchs2008}. State-of-the-art diamond turning can produce low-NA optical surfaces 50 mm in diameter with peak-to-valley surface deviations of 150 nm \cite{Kim2009} and local surface roughness below 0.4 nm \cite{Marsh2005}. This sub-wavelength precision has been utilized in quantum optics applications such as laser mode converters \cite{Blough1997, Shen2013b, Mishra2007}, monolithic microcavities \cite{Brieussel2016}, and other resonators \cite{Klaassen2007}.

%axes figure

\begin{figure}[t]
\centerline{\includegraphics[width=135mm]{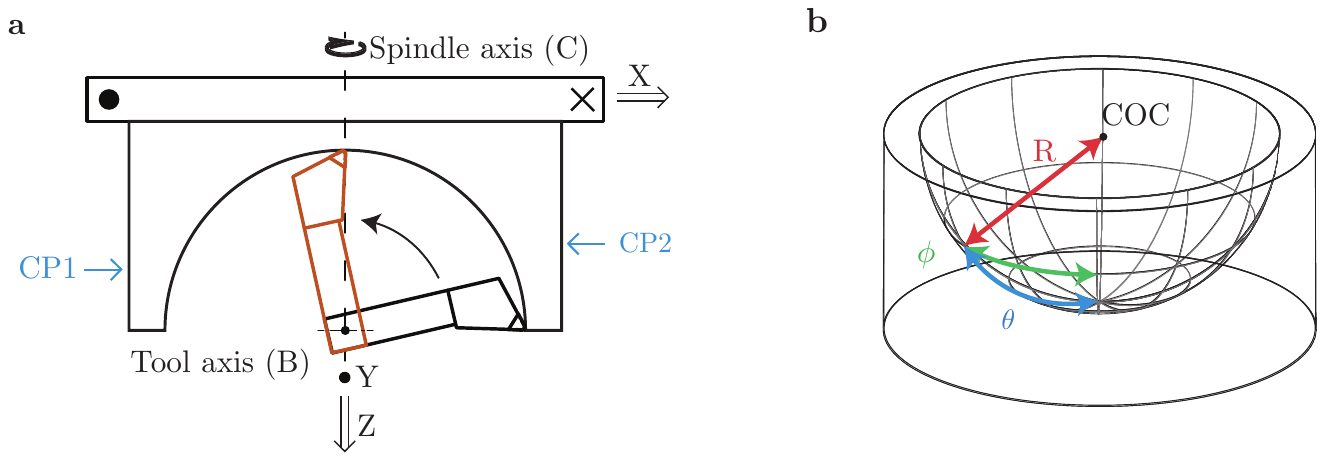}} 
\caption{(a) Top view of the diamond turning tool path showing the two linear lathe axes X and Z, two rotational axes B (tool) and C (spindle), and tool height axis Y (out of plane). The cutting tool is shown in its initial (black) and final (orange) position. Also shown are the counter-clockwise (CP1) and clockwise (CP2) contact points used for fine-calibrating the rotational axes centers as described in the text. (b) Hemisphere with surface coordinate system: radial distance (R), zenithal angle ($\phi$) and azimuthal angle ($\theta$).}
\label{fig:axes}
\end{figure}

%model figure

\begin{figure}[t!]
\centerline{\includegraphics[width=135mm]{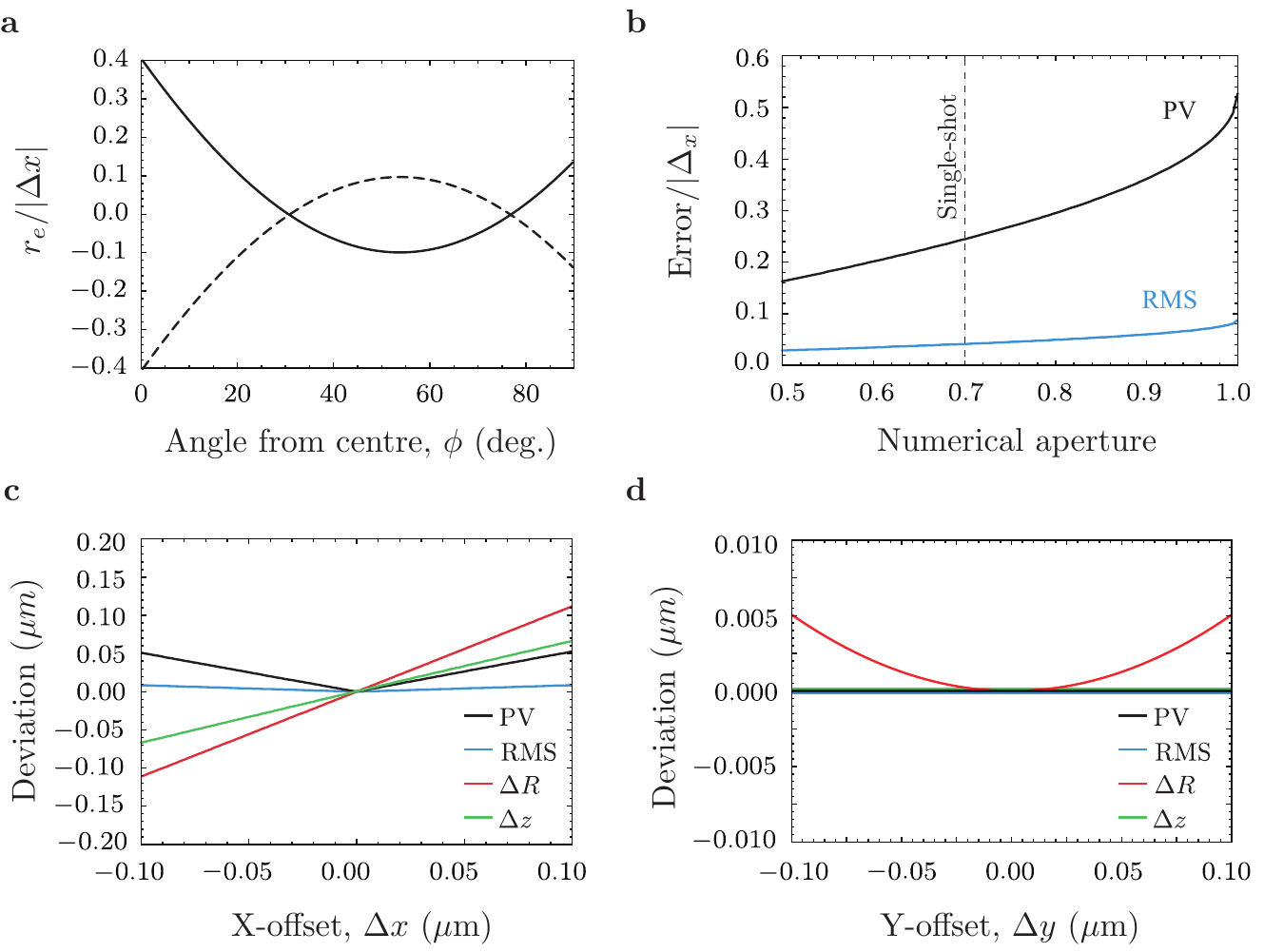}} 
\caption{Consequences of calibration errors. (a) Radial error $r_e$ from a positive (solid) and negative (dashed) X-axis offset as a function of zenithal angle $\phi$. (b) The PV (black) and RMS (blue) aggregate error of a surface with error profile $r_e$ measured over a central region as a function of the aperture. (c and d) Form errors and fit parameters from an X-axis (c) and a Y-axis (d) offset. The PV error (black) and RMS error (blue) are proportional to the X-offset but insensitive to a Y-offset. In each case a spherical fit to the cut surface differs from the tool arc by a change in the RoC by $\Delta R$ (red) and CoC position shift $\Delta z$ (green).}
\label{fig:model}
\end{figure}

We diamond turn hemispheres on a CNC nano-lathe, the Nanotech 250UPL from Moore Precision Tools, the configuration of which is shown in Fig.~\ref{fig:axes}a. The lathe has four precision controlled  degrees of freedom: an aerostatic spindle that rotates the workpiece at 2,000 RPM around the spindle axis (C); two perpendicular linear axes (X and Z) for positioning the cutting tool and spindle on fully-constrained oil hydrostatic bearings; and an additional axis (B) that rotates the tool post on a groove-compensated air bearing. The final degree of freedom (Y) is the height of the cutting edge compared to the center of the spindle, over which we have limited manual control. The X, Z, C and B axes are interferometrically stabilized with control resolution 1 nm (X, Z), 1 arc second (C, B) and feedback resolution 1 pm (X, Z), 0.01 arc seconds (C, B). The B axis rotates with radial and axial displacement error less than 50 nm over the full 360$^{\circ}$, of which we require 90$^{\circ}$. The Y axis is manually adjusted with $\mu$m resolution and not actively stabilized. The spherical coordinate system for the cut surface is shown in Fig.~\ref{fig:axes}(b) with zenithal and azimuthal angles $\phi$ and $\theta$ related to the rotational axes B and C of the lathe respectively. 

The hemispheres are cut from a cylindrical aluminum 6061 substrate with outside diameter 30 mm and height 13.5 mm. Because of their two-fold rotational symmetry, hemispheres may be cut by a single rotation of the B axis through $90^\circ$ from the edge to the center of the hemisphere, repeated at increasing depths by iterating the tool post towards the spindle along the Z axis. This technique uses a single point on the diamond tool edge to cut across the entire surface with two key consequences: the surface form is insensitive to the exact shape of the cutting tool edge, and the relative cutting force direction and magnitude is constant over the cut \cite{Drescher1990}. The mirror radius of curvature (RoC) is determined by the distance from the most remote point on the cutting tool edge to the B axis center of rotation ($R_c$) and by the offset between the two rotating axes. We determine $R_c$ by measuring the tool edge position as a function of B-axis angle with an optical microscope fixed above the lathe. In the limit of many measurements this approach should determine $R_c$ to within $1$~$\mu$m, although in practice we performed this calibration only to a measurement uncertainty of $10$~$\mu$m.

In contrast to the finely polished mirrors used for high-finesse cavities, SPDT produces a finished surface with a turning spiral that tracks the passage of the tool across the part. The groove spacing is $s =  2 \pi \frac{\omega_B}{\omega_C} R_c$ where $\omega_B$ and $\omega_C$ are the rotational speed of the tool and spindle axes respectively. Assuming the mirror radius $R(\theta, \phi) \approx R_c$ is much larger than the tool edge radius $r$, the min-max height of the grooves $h$ is $h = r-\sqrt{r^2 - (\frac{s}{2})^2}$. Together these parameters determine a minimum cut time 
\begin{equation}\label{eqn:cut_time}
T_{min} = \frac{\pi R_c}{4 \omega_C \sqrt{h_\mathrm{max}(2r - h_\mathrm{max})}}
\end{equation}
for a surface with maximum groove height $h_{max}$. The time to cut a hemisphere with sub nm grooves in our case ($R_c = 12.6$ mm, $r = 0.6$ mm, $\omega_C = 2000$ rev/min) is 5 minutes.

It is feasible to cut mirrors larger than those demonstrated here (our lathe is capable of cutting mirrors with RoC up to $250$~mm) so long as the temperature of the lathe environment is stable over the duration of the cut. Temperature variations of just $0.5^\circ$~K are enough to significantly degrade form accuracy due to the thermal response of aluminium. At the other end of the scale, the mirror size cannot be made smaller than available diamond tool heads and so it would prove challenging to manufacture precise hemispheres with RoC $<1$~mm by this technique.

\subsection*{Interferometric axes calibration} 

High-NA optics manufactured with SPDT are critically sensitive to the relative rest positions of the lathe axes. Any offset between the spindle rotation axis and the tool path symmetry axis causes a $\theta$ symmetric, $\phi$ dependent deviation that scales poorly with NA and quickly comes to dominate the error profile. When the X-axis is configured such that the rotational centers of the C and B axes are aligned, the B rotation cut describes a perfect circle with constant RoC. However when the C-axis center is slightly past (before) the spindle center C the tool will cut a slightly larger (smaller) radius and produce a distinctive radial error profile
\begin{equation}
r_e = -\Delta x[ (\frac{3\pi}{2}-4)\cos{\phi} + \sin{\phi} + (2-\pi)] \,.
\end{equation}
with a single minimum (maximum) at the zenithal angle $\phi = 54.5^\circ$. 

Typical X-axis offset error profiles are illustrated in Fig.~\ref{fig:model}a where the radial deviation from a perfect hemisphere is given as a function of the zenithal angle $\phi$ from the mirror center. Fig.~\ref{fig:model}b shows how the aggregate form error from an  x-offset $\Delta x$ scales with the NA of the hemisphere. As a consequence of this misalignment the CoC position along the optical axis of the hemisphere is  corrected to match the new best-fit RoC, Fig.~\ref{fig:model}c shows how the RoC, CoC, RMS error and PV error scale with $\Delta x$. Measured over the full hemisphere the PV and RMS form errors scale like $E_\mathrm{PV} = \frac{\Delta x}{2}$, $E_\mathrm{RMS} = \frac{\Delta x}{11}$. Y-offsets have a negligible impact on the form error, but do introduce a small adjustment to the CoC position as shown in Fig.~\ref{fig:model}d as a function of offset $\Delta y$ and a central defect of width $\Delta y$. A derivation of the X and Y-offset error profiles, and a schematic of the relevant coordinates are included as supplementary material.

To align the two rotational axes of the lathe, shallow calibration cuts are made on the outer surface of the aluminum substrate by moving the diamond cutting tool perpendicularly into the surface. The cuts are performed on alternating sides of the substrate, at positions CP1 and CP2 shown in Fig.~\ref{fig:axes}a, using alternating spindle rotation directions. Between each of the cuts, the B axis is rotated by 180$^\circ$ and $\Delta x$ can be inferred by measuring the depth of subsequent cuts.

The depth measurement is performed using a white-light interferometer that is mounted adjacent to the tool post of the lathe. The instrument consists of a 20X Nikon Mirau interferometry objective mounted to a fixed focal-length video microscope with in-line white-light illumination. A schematic of the interferometry scheme is provided as supplementary material. This in-situ surface metrology provides a depth resolution of 0.4 nm by fitting a complete interference fringe. To reduce the duration of the calibration procedure, however, we used a quick estimate of the relative cut depths by observing the fringe pattern while varying the position of the interferometer. This method provided a resolution of approximately 10 nm, which is sufficient to achieve our target form accuracy.

\section*{Results and discussion}

\subsection*{Surface profiles}

\begin{figure}[t]
\centerline{\includegraphics[width=80mm]{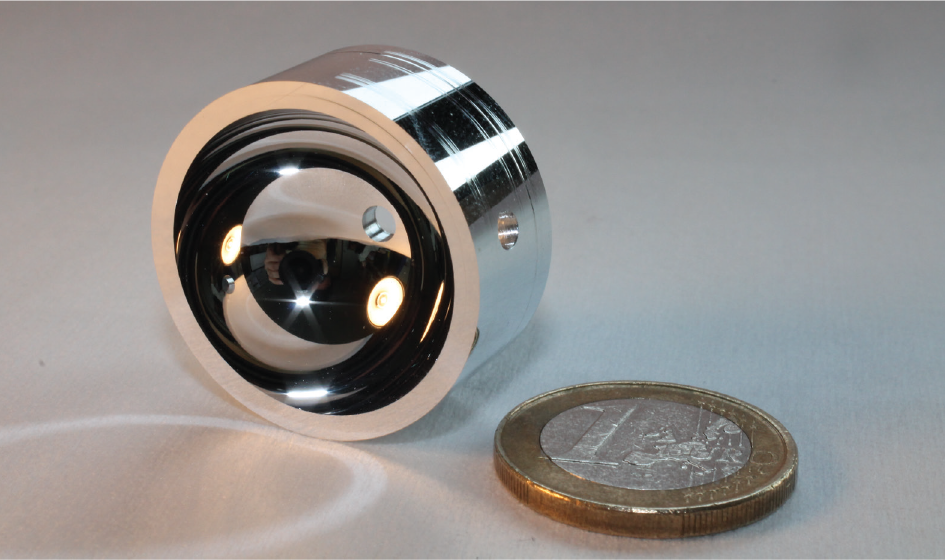}} 
\caption{Hemispherical mirror, part E, with one Euro coin for scale. The mirror radius of curvature is $12.578(1)$~mm. This mirror was turned from a substrate with pre-drilled beam ports, which are visible in the photo. Also visible on the exterior surface of the mirror are shallow grooves cut during the lathe calibration process.}
\label{fig:photo}
\end{figure}

A finished hemisphere cut with the above technique is shown in Fig.~\ref{fig:photo}. Two complementary interferometric measurements are used to verify the finished surface profile. The primary measurement is a large area surface profile by optical interferometry with a ZYGO interferometer. This measures the mirror surface against the optical wavefront produced by a reference sphere with PV error 20 nm. The size of the available reference sphere restricts these measurements to a NA of 0.7 (half angle $44.4^\circ$). To profile a complete mirror surface it is necessary to `stitch' together separate scans that cover the entire surface, which requires at least eight scans of NA 0.7 \cite{Kershner1939}. However a single NA 0.7 interferogram taken at 45$^\circ$ to the optical axis of the hemisphere is sufficient on its own to measure the complete zenithal profile of an axially symmetric part, which is typically the largest component of the error profile of the hemispheres.

%axes figure

%Table environment contains table and figure to force both to same page
\begin{table*}[t!]\centering
%\setlength\tabcolsep{6pt}
%\setlength\extrarowheight{4pt}

%Table with booktabs
\begin{tabular}{@{}lcrrcrrcrr@{}}\toprule

&\phantom{\hspace{15mm}} &\multicolumn{2}{c}{Total} & \phantom{\hspace{15mm}}& \multicolumn{2}{c}{Zenithal} & \phantom{\hspace{15mm}} & \multicolumn{2}{c}{Azimuthal}  \\
\cmidrule{3-4} \cmidrule{6-7} \cmidrule{9-10}
Part&&PV & RMS &&  PV & RMS && PV & RMS\\ \midrule

A && 146.2 & 22.1 && 89.4 & 19.5 && 91.9 & 10.4 \\
B && 87.8 & 13.5 && 59.1& 11.4&& 68.6& 7.1  \\
C &&144.7&18.3&&  52.7& 14.1&&  114.2& 11.6 \\
C* && 215.0&27.4&&  47.5& 14.2&&  207.7& 23.4 \\
D && 317.3&51.2&&  219.4& 44.7&&  193.2& 23.0 \\
E && 116.5&18.1&&  76.2& 15.8&&  73.8& 8.7 \\

\bottomrule
\end{tabular}
\caption{Aggregate form errors of five diamond-turned hemispheres, all quantities in nm. The total  errors are the typical (RMS) and maximal (PV) radial deviation from an ideal hemisphere over the complete mirror surface. The zenithal profile (orange trace in Fig. \ref{FullHems_fig}) is used to separate zenithal and azimuthal contributions to the form error. Part C* is a repeat measurement of part C after bake-out for ultra-high vacuum. Part E was turned on a substrate with pre-drilled beam ports. See text for details.}
\label{table}
\end{table*}

\begin{figure}[t!]
%Figure 10B and 11C

\centerline{\includegraphics[width=150 mm]{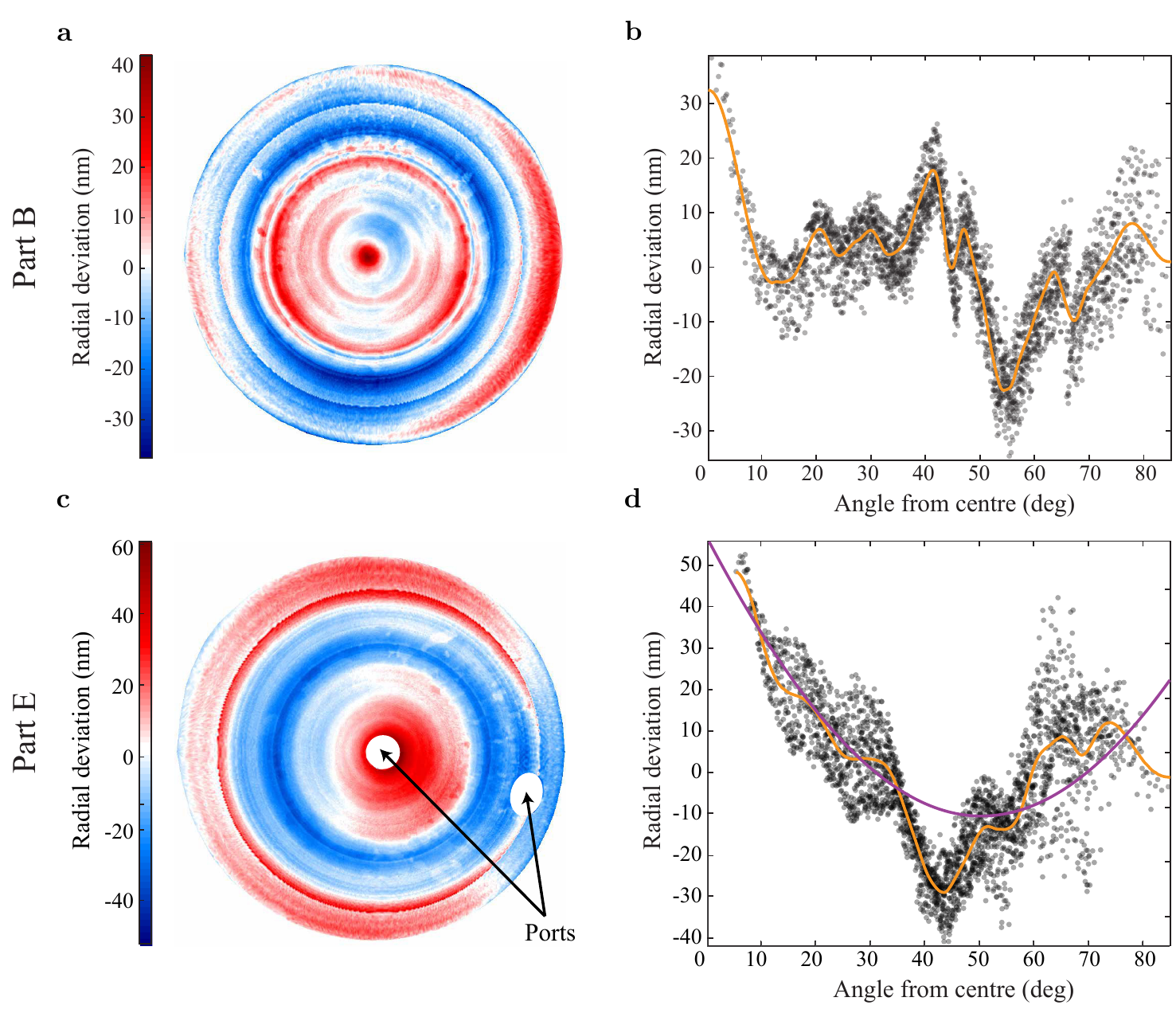}} 
\captionof{figure}{Surface detail of two hemispheric mirrors (Tab.~\ref{table} lists the aggregate errors of all measured mirrors). (a) Radial error profile of part B as an azimuthal equidistant projection. The complete surface is reconstructed from stitched interferograms. (b) The radial error profile of part B as a function of zenithal angle ($\phi$) from the axis of rotation (with reduced resolution, every 40th pixel of the stitched interferogram is plotted). A Savitsky-Golay filter of the radial error (orange line) separates the zenithal and azimuthal components. (c) Radial error profile of part E by the same method. The two beam ports are shown. (d) The radial error profile of part E as a function of zenithal angle including Savitsky-Golay filter (orange). The X-offset model $e_r$ (purple) from Fig.~\ref{fig:model}a shows the contribution of the residual X-offset to the measured profile, with an implied offset of $\Delta x = 150 \pm 10$ nm. }
\label{FullHems_fig}
\end{figure}

The partial profiles measured by reference-sphere interferometry are in agreement with stitched surface profiles taken by Meopta - optika. These complete surface profiles are stitched from a hundred small interferograms taken against a reference flat. Two complete surface profiles are shown in Fig.~\ref{FullHems_fig}a and c. The circular density map is an azimuthal equidistant projection of the hemisphere onto the plane. Because the mirrors retain a degree of azimuthal symmetry it is illuminating to plot the same data as a function of zenithal angle $\phi$ to see the mean zenithal radius dependence, as shown in Fig.~\ref{FullHems_fig}b and d. For clarity these zenithal plots are shown with reduced resolution, every 40th pixel of the stitched interferograms is plotted. The scatter plot density scales with the value of the differential area element $dA/d\phi \approx \sin (\phi)$. A 3rd-order, 200 point Savitsky-Golay filter of the scatter plot separates the azimuthal and zenithal components of the form error. The zenithal radius function reconstructed in this way is consistent with contact probe measurements taken along an arc through the mirror center.

In Table~\ref{table} we compare five hemispheres. The results demonstrate the consistent fabrication of hemispheres with RMS error under 25~nm and as good as 14~nm. For each mirror the results are separated into zenithal and azimuthal components by the method described above. Parts A, B, C and D are consecutive attempts to cut the same hemisphere. Of these, part D is an outlier that is included for the sake of completeness. Its error profile indicates an abrupt  and unusual shift in the position of the lathe during the cut that remains unexplained. The mirrors were cut with $R_c$ in the range $12.55(1)$ - $12.60(1)$~mm, and the measured RoC of each mirror is within the $10$~$\mu$m precision of our $R_c$ calibration. Each profile features a peak at $0^\circ$ as the cutting force vanishes towards the center of the spindle that may be compensated by a progressive adjustment of the Z position over the final few degrees of the cut. 

By fitting the X-offset model (Fig. ~\ref{fig:model}a) to the measured surface profile we can infer the residual X-offset that remains after our on-lathe calibration technique. Part B, the best part yet measured, has a residual offset of $29 \pm 3$ nm, which accounts for only 20\% of the total form error and which is largely spurious fitting to unrelated temperature fluctuations. However the typical residual offset is larger, as illustrated by the X-offset model fit to part E plotted in Fig.~\ref{FullHems_fig}d which implies a residual offset of $150 \pm 10$ nm. Although this is the largest measured residual offset (excluding part D), the typical offset is considerably larger than the tolerance of the calibration process. The source of this shift, a gradual tilt due to a deflating vibration isolation airbag, has been identified and removed so that subsequent hemispheres should be more consistently similar to part B.  

\subsection*{Experiment capability}

In proposed QED experiments with hemispherical mirrors \cite{Hetet2010}, a neutral or ionized atom would be confined to a region much smaller than the transition wavelength at the hemisphere CoC by either an optical or electromagnetic trap. In either case, UHV conditions are required to isolate the trapped atom from background collisions, and optical access from several directions is required for cooling. Two further measurements were made to test the suitability of these hemispheres for such experiments. First, part C was measured before and after bake-out in vacuum as preparation for UHV experiments. The bake-out introduces an increase in both PV and RMS form errors (C* in Table~\ref{table}) due to the relaxation of material strain as the temperature is increased to $200^\circ$, held for 2 hours and then reduced to room temperature. In contrast to the cutting errors, this distortion is not rotationally symmetric. An analysis of the distortion is included as supplementary material.

Second, part E was cut to demonstrate the feasibility of diamond turning hemispheres from substrates with pre-drilled holes for beam ports. Proposed measurements with these hemispheres require on- and off-axis laser access to the mirror CoC through the mirror. Two beam ports with 3 mm diameter were drilled through the substrate of part E before the surface was lathed, one along the rotational axis and one at $\phi_\mathrm{port} = 62^\circ$ from the center. The two ports are visible in the complete surface profile of part E in Fig.~\ref{FullHems_fig}. The ports introduce cutting force variations visible at the zenithal angle of the port, and this degrades the form accuracy slightly as seen by comparing part E (RMS error 18 nm) to parts A, B and C (22, 14 and 18 nm).

%QED figure
\begin{figure}[t!]
\centerline{\includegraphics[width=100mm]{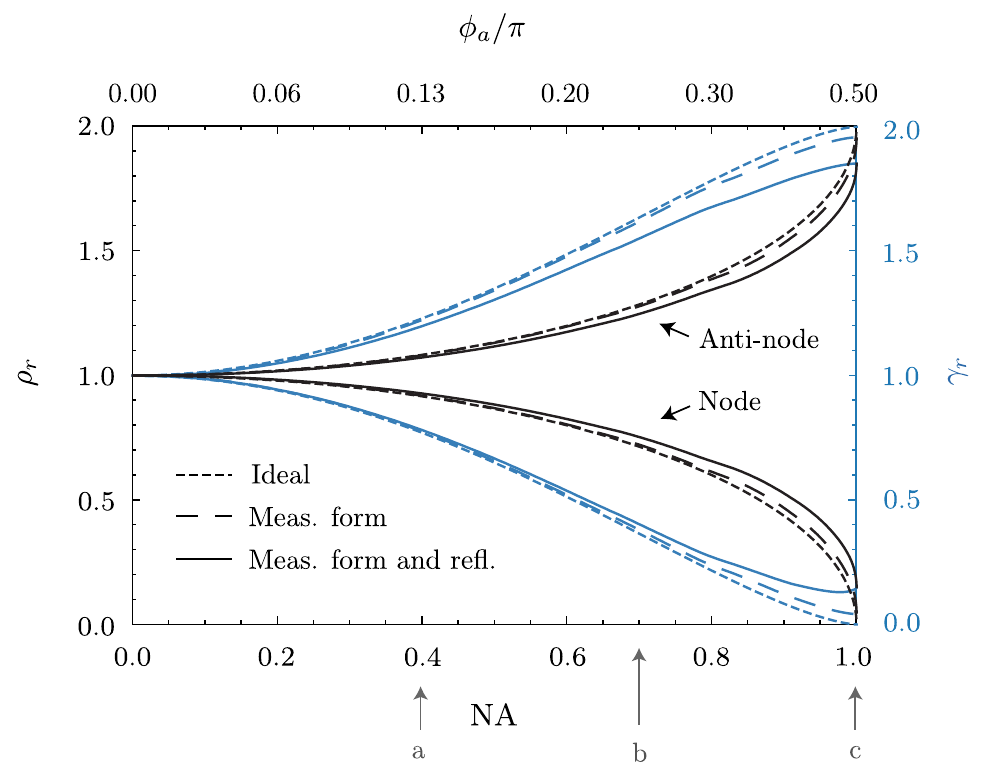}} 
\caption{Modeled performance of part B in QED experiments. (black) Relative vacuum mode density $\rho_r$ at the CoC of a spherical mirror as a function of NA (top axis shows the corresponding half-aperture in radians). The CoC is a node (anti-node) of the vacuum mode density at wavelength $\lambda$ for $R = n\frac{\lambda}{2}$ ($R = n\frac{\lambda}{2} + \frac{\lambda}{4})$. (blue) The relative spontaneous emission rate $\gamma_r$ of a linear dipole at the node/anti-node emitting at wavelength $\lambda$. Dotted lines are the result of an ideal spherical mirror \cite{Hetet2010}, dashed lines are the result of a perfectly reflective sphere with the surface of part B (to a given NA), and solid lines are the same result including the known reflectivity of aluminium at $\lambda$=493~nm. Also shown are the NAs of (a) previous attempts to measure such an effect \cite{Eschner2001} (b) the best available diffraction-limited focusing optics and (c) this work. Considering its NA, reflectivity and form; part B is modeled to reduce or enhance both $\rho$ and $\gamma$ by 88\%.}
\label{fig:QED}
\end{figure}

\subsection*{QED performance}

Part of our motivation for this work is to develop free-space QED systems. Armed with detailed metrology of our surface quality we can now calculate how our mirror would modify the emission of a dipole placed at the CoC. Fig.~\ref{fig:QED} shows how the relative vacuum mode density ($\rho_r = \rho/\rho_0$) and relative spontaneous decay rate ($\gamma_r = \gamma/\gamma_0$) are altered by retro-reflection with spherical optics according to the model in Ref.~\cite{Hetet2010}. Whether this QED effect enhances or suppresses the vacuum mode density depends on the RoC. We do not, however, need to machine a hemisphere with an RoC calibrated to within a fraction of a wavelength. Like spheres, hemispheres expand uniformly about the centre of curvature. The coefficient of thermal expansion of aluminium 6061 at room temperature is $23.5 \times 10^{-6}$ K$^{-1}$ \cite{Totten2003}. A temperature change of $.42^\circ$~K is enough to switch between enhancement and suppression of spontaneous emission at the focus of our mirror. So provided we achieve suitable form accuracy, the average radius can be actively tuned. Resistively heating the mirror under the single-shot, white-light ZYGO interferometer shows tunability over the desired range, and no indication of surface distortion due to thermal expansion.

The key to a strong QED effect is maximizing the NA. The experiment of Ref.~\cite{Eschner2001} used a multi-element lens and planar mirror combination with NA=0.4 (arrow \textit{a} in Fig.~\ref{fig:QED}). According to the model, the density of modes is reduced by 8\% giving suppression of spontaneous emission by 24\% at this NA. The highest NA lens one could realistically hope to use is about 0.7, limited by the geometry of the experiment. This point is shown by arrow \textit{b} and reduces the density of modes by 30\% and decay rate by by 65\%. Our work has shown the machining of a spherical mirror with NA=0.996 (arrow \textit{c}). The model allows us to include the reflectivity of aluminium (0.92 at $\lambda = 493$~nm \cite{Mondolfo1979, Smith1997}) and the measured form error (Fig.~\ref{FullHems_fig}). Using the data for part B, we predict suppression of the density of modes and spontaneous emission rate of 88\%. This figure is dominated by the reflectivity of the surface. A perfectly reflective surface with the same form arrow would give 96\% suppression of mode density and emission. High-reflectivity surface coatings are one approach to improving reflectivity. Metallic coatings can improve on the reflectivity of aluminium at some wavelengths, and can be deposited over high-aperture surfaces without introducing substantial surface distortion. However, multi-layer dielectric coatings, which can be very close to perfectly reflective on low-aperture mirrors, require a higher degree of uniformity. Typical sputtering techniques are too directional to produce a high-reflectivity dielectric coating over a hemisphere.

\section*{Conclusions}

We have demonstrated the fabrication of high-NA hemispheric optics for quantum optics experiments and metrology by single-point diamond turning. Our best sample achieves NA = 0.996, PV error 88 nm and RMS error 14 nm. These deviations from spherical are within a factor of five of the roundest manufactured spheres \cite{Everitt2011, Andreas2011a}, even without polishing. This accuracy meets the requirements for experiments that seek strong free-space atom-light coupling. An ideal hemispheric mirror can reduce the vacuum mode density, and therefore the atomic fluorescence rate of an atom at the mirror CoC, to zero \cite{Hetet2010}. Our best mirror (part B) is close enough to spherical to reduce or enhance the vacuum mode density by 96\%. With this degree of reservoir engineering it is possible to investigate QED effects with free space optical modes, and tune the emission spectra and spatial mode of emitters at the mirror CoC.
Four consecutively cut hemispheres demonstrate that this technique reliably produces surfaces with RMS error below 25 nm and further measurements show the suitability of these mirrors for proposed experiments, with the capacity to turn a comparable surface over a substrate with pre-drilled beam ports and minimal disturbance by bake-out for use in UHV. A crucial part of our setup was the development of an on-lathe white-light interferometer that was applied to measure a series of calibration cuts made immediately before the precision machining of a mirror. This calibration process has made possible the production of hemispheric reference surfaces with precision previously feasible only for reference flats and could be applied similarly for the diamond turning of high-NA parabolic and axially asymmetric reflectors and optics.

% Bibliography
\bibliographystyle{unsrt}
\bibliography{Mendeley}

\section*{Acknowledgments}

Our work was funded by the Australian Research Council (ARC) (CE1101027, FL150100019) and received financial support from the Institut f\"{u}r Quanteninformation GmbH and the Austrian Science Fund (FWF) through projects P23022 (SINPHONIA) and F4001 (SFB FoQus). We gratefully acknowledge the contribution of W. Shihua and colleagues at the National Metrology Centre, A*STAR in Singapore. Without their expertise and the use of their high-NA optical interferometer and contact-probe measurements the first stages of this work would not have been possible. We also acknowledge the scientific advice of Prof. Rainer Blatt at the Institut f\"{u}r Experimentalphysik, Universit\"{a}t Innsbruck.

\section*{Author contributions}
D.H. and G.C. fabricated the mirrors. G.C. designed the on-lathe white-light interferometer. D.H. conducted surface interferometry of the mirrors. G.A. and Y.C. performed the bake-out test and advised on experimental applications. F.F. advised on diamond turning and nanofabrication. B.B. and P.L. supervised the manufacture and characterization of the mirrors. D.H. prepared the manuscript with input from all authors.

\section*{Additional information}

\subsection*{Competing financial interests}

The author(s) declare no competing financial interests.

\subsection*{Supplementary material}

Supplementary material is available online at https://www.nature.com/articles/s41598-017-18637-8 

\subsection*{Data availability}
The datasets generated during and/or analyzed during the current study are available from the corresponding author on reasonable request.

\end{document}